\documentclass{mem}
\usepackage{natbib}\usepackage{txfonts}\usepackage{balance}
\usepackage{graphicx}
\usepackage[a4paper,breaklinks,dvipdfm]{hyperref}
\idline{75}{282}
\begin{document}
\def\teff{$T\rm_{eff }$}
\def\kms{$\mathrm {km s}^{-1}$}

\title{
Angular momentum and disk evolution in very low mass systems
}

   \subtitle{}

\author{
Alexander Scholz\inst{1,2} 
          }

  \offprints{A. Scholz}

\institute{
School of Cosmic Physics, Dublin Institute for Advanced Studies, 31 Fitzwilliam Place, 
Dublin 2, Ireland
\and
School of Physics \& Astronomy, University of St. Andrews, St. Andrews, KY16 9SS, United Kingdom
\email{as110@st-andrews.ac.uk}
}

\authorrunning{Scholz}

\titlerunning{Angular momentum and disks}

\abstract{
This review summarises recent observational results regarding the evolution of angular
momentum and disks in brown dwarfs. The observations clearly show that brown dwarfs beyond
ages of 10\,Myr are exclusively fast rotators and do not spin down with age. This suggests
that rotational braking by magnetic winds becomes very inefficient or ceases to work in the 
substellar regime. There is, however, some evidence for braking by disks during the first
few Myrs in the evolution, similar to stars. Brown dwarf disks turn out to be scaled down
versions of circumstellar disks, with dust settling, grain growth, and in some cases cleared
out inner regions. The global disk properties roughly scale with central object mass. The 
evolutionary timescales in substellar disks are entirely consistent with what is found for 
stars, which may be challenging to understand. Given these findings, it is likely that 
brown dwarfs are able to form miniature planetary systems.
}

\maketitle{}

\section{Introduction}

Angular momentum and disk evolution are rarely combined in a review, but these
two topics are fundamentally related to each other. The disk controls 
the early angular momentum evolution of stars. Many aspects in the evolution of disks, 
including those relevant for planet formation, are related to the angular 
momentum transport in the disks and thus have an indirect connection to the 
rotation of the central objects. It is now established that the rotation 
of young stars is regulated by interaction with their disks ('disk braking').

This emphasis of this review, however, are observational results. Apart from the 
disk braking mentioned above, there are very few direct observational links 
between rotation and disks. This is even more so in the substellar regime, 
where studies are generally hampered by the faintness of the sources. I will 
therefore treat these two subjects mostly separately.

I will focus on brown dwarfs, although some of 
the objects discussed below may turn out to be stars very close to the Hydrogen 
burning limit. Brown dwarfs are intriguing systems because they allow us to probe 
physical processes as a function of the object mass. Therefore, I will introduce
the two main sections with a brief summary of the status of our understanding
in solar-mass and low-mass stars. This can be used as a reference point to
test the brown dwarf regime. 

\section{Angular momentum evolution}
\label{s2}

The rotational evolution of low-mass stars can be divided in two distinct
phases, which can also be used as a framework for interpreting the 
findings for brown dwarfs: \cite[see reviews by][]{2007prpl.conf..297H,2009AIPC.1094...61S}

(1) During the first few million years, the rotation is strongly 
regulated in the sense that angular momentum is not conserved and the period,
averaged over a large sample, stays largely constant. This regulation is 
presumably due to some interaction between star and disk, in the following 
called disk braking.

(2) After the disks have disappeared, the rotational evolution is determined
by the spinup due to pre-main sequence and, on longer timescales, by angular
momentum losses due to magnetically driven stellar winds. This
causes the spindown of solar-mass stars to rotation periods of weeks.

Rotation periods measured from inhomogenuously distributed surface features
co-rotating with the objects are the ideal observational diagnostics of stellar 
and substellar rotation. The surface features can be magnetically induced spots
or, as in ultracool dwarfs, patchy clouds. Although I will focus on periods, 
an assessment of the rotational evolution is also possible through measurements 
of spectroscopically determined rotational velocities which provide important
complementary information.

\begin{figure*}[t!]
\resizebox{\hsize}{!}{\includegraphics[clip=true]{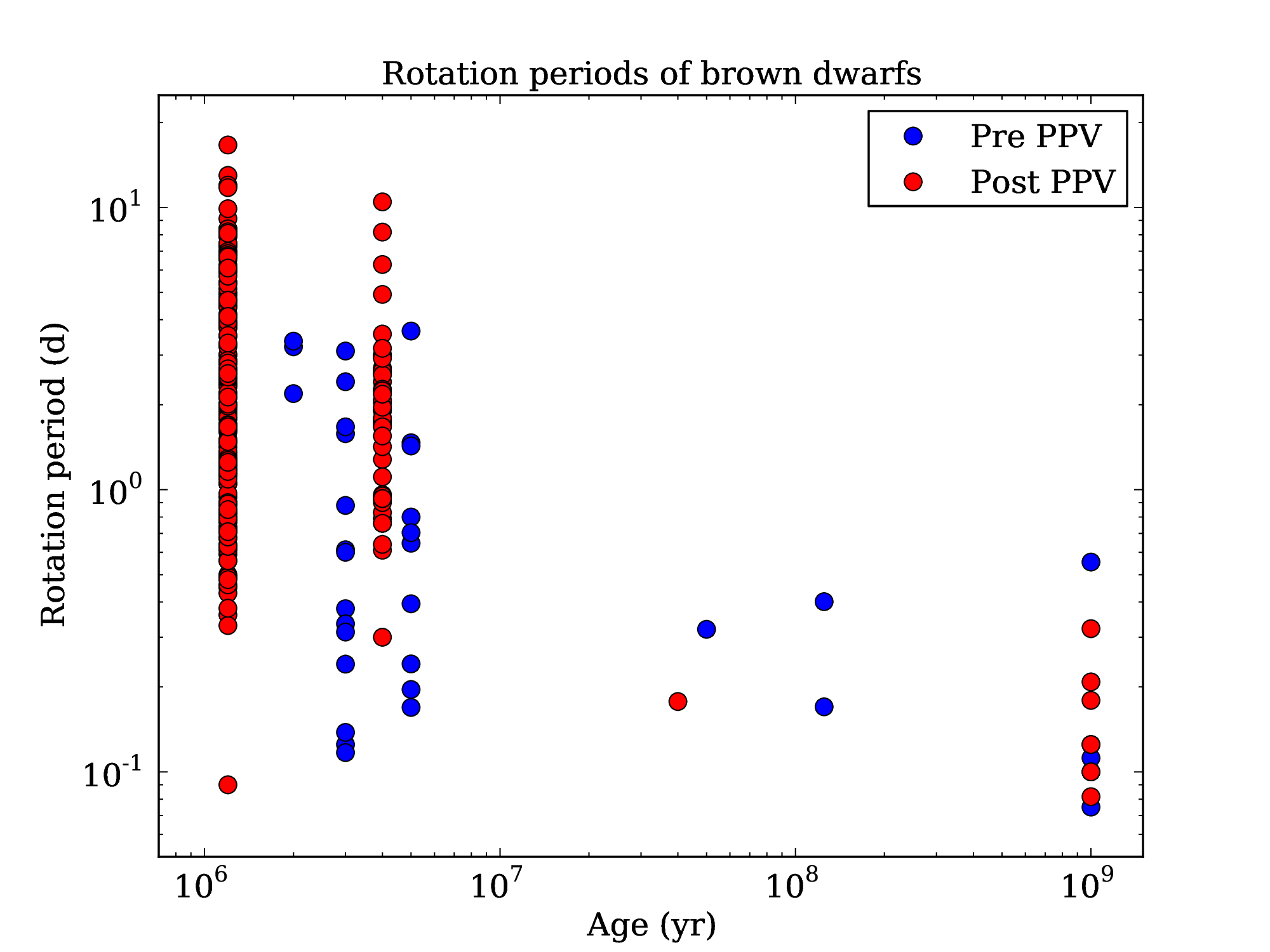}}
\caption{\footnotesize Rotation periods of brown dwarfs. Periods published before 
and after the Protostars \& Planets V review by \citet{2007prpl.conf..297H} are 
shown in blue and red, respectively. Periods for ultracool field dwarfs with unknown 
age are plotted at 1\,Gyr. The figure contains datapoints from: 
\citet{1997MNRAS.286L..17M,1999AJ....118.1814T,2001A&A...367..218B,2002MNRAS.332..361C,2003A&A...408..663Z,2003ApJ...594..971J,2004A&A...419..249S,2004A&A...421..259S,2004A&A...424..857C,2005A&A...429.1007S} 
for 'Pre PPV' and 
\citet{2006MNRAS.367.1735K,2007ApJ...668L.163L,2009A&A...502..883R,2009ApJ...701.1534A,2009MNRAS.400.1548S,2010ApJS..191..389C,2012ApJ...750..105R,2013ApJ...767...61G,2013ApJ...767..173H,2013arXiv1304.0481G}
for 'Post PPV'.}
\label{f1}
\end{figure*}

Six years ago, only about 30 rotation periods for brown dwarfs were known, see
Fig. 6 in \citet{2007prpl.conf..297H}. In the last years, the period database 
has grown significantly, but it is still sparsely populated,
particularly at ages $>10$\,Myr. More datapoints are needed at all ages to provide 
further constraints and to improve the statistics.
In Fig. \ref{f1} we show the currently known
periods for brown dwarfs, to illustrate the main trends and the currently existing 
samples. While the periods for very young objects (ages $<10$\,Myr) show a wide range,
from several hours up to 20 days, the periods for older objects are all shorter
than 1 day.

\subsection{The evolved objects}

The evolved objects in this plot with ages $>10$\,Myr do not possess disks.
Their rotational evolution is only affected by contraction and wind braking.
As mentioned above, all periods for evolved brown dwarfs are shorter than 1 day. 
This is confirmed by the available spectroscopic data. The lower limit of the 
projected rotational velocities for ultracool dwarfs increases from mid M to
L dwarfs, with the effect that all L and T dwarfs have $v\sin{i}$ larger than
7kms\,$^{-1}$, corresponding to periods shorter than 17\,h
\citep{2008ApJ...684.1390R,2010ApJ...710..924R,2010ApJ...723..684B,2012ApJ...750...79K}.

Thus, the observations show very convincingly that all evolved brown dwarfs
are fast rotators. This can only mean that wind braking becomes extremely
inefficient in the substellar regime or even ceases to work. This continues
a trend that has been noticed in the stellar regime by several groups: the
spindown timescale is a strong function of mass and increase towards
lower mass objects. Stars with 0.3$\,M_{\odot}$ need 0.5\,Gyr to spin down
\citep{2011MNRAS.413.2595S}.\footnote{This assumes an exponential spindown law
with $P \propto \exp{(t/\tau)}$.} For 0.1$\,M_{\odot}$ this timescale increases to several 
Gyr \citep{2011ApJ...727...56I}. The observational data suggests that the 
rotational 
braking becomes gradually less efficient towards lower masses, until it 
essentially shuts down for brown dwarfs. Brown dwarfs might still spin down, 
but it could take more than the current age of the Universe. A slowly rotating 
brown dwarf would constitute a fossil object from the beginning of the Galaxy.

The transition from slowly rotating main sequence stars like the Sun to
fast rotating brown dwarfs is remarkable, because it puts brown dwarfs in the
regime of giant plants when their long-term rotational evolution is considered.
Similar to giant planets and in stark contrast to stars, brown dwarfs do not
spin down as they age and continue to be fast rotators. This also means that,
in contrast to stars, coeval brown dwarfs with similar masses do not necessarily
have similar rotation periods.

It is not clear yet what causes this breakdown of
wind braking in the very low mass regime. This fact has often be interpreted as a sign 
of a change in the
magnetic field generating dynamo \citep[e.g.][]{2004PhDT.........4S}, but this is 
unlikely
to be the only cause \citep[see][]{2009MNRAS.400.1548S}. Changes
in magnetic field topology could also play a role \citep{2010MNRAS.407.2269M}. 
Furthermore, as argued by \citet{2012ApJ...746...43R}, adopting a modified
spindown law with a retained dependence on the radius could explain 
this transition without any further changes in the physics. Independent
observations of magnetic field indicators as well as further critical 
evaluation of the theoretical frameworks are important steps to answer this
question.

\subsection{The young objects}
\label{s23}

The rotation periods for young objects provide information
on the initial angular momentum content of brown dwarfs. 
The periods exhibit a large scatter, from fractions of a day up to 10 days, with 
a tail to even longer periods. While the total spread in periods is similar to 
more massive stars, the period distribution is not. In all these samples there 
is a consistent trend of 'faster rotators lying towards lower masses' 
\citep{2009A&A...502..883R}. In the ONC, the median period drops from 5\,d 
for $M>0.4\,M_{\odot}$ to 2.6\,d for VLM stars and to 2\,d for BDs. Thus, there 
is a clear mass dependence in the rotation periods at very young ages. As noted 
by \citet[][their Fig. 10]{2010ApJS..191..389C} and earlier by 
\citet{2001ApJ...554L.197H}, this period-mass trend is consistent with specific 
angular momentum being independent of object mass. This equipartition of angular 
momentum is an interesting outcome of the star formation process and could 
provide a useful constraint on theories for collapse and fragmentation of 
clouds.

A controversial aspect of the brown dwarf periods in star forming regions is 
their lower limit. The ‘breakup limit’, where centrifugal forces are in balance 
with gravity, is between 3 and 5\,h at these young ages. 
\citet{2003A&A...408..663Z}, \citet{2004A&A...424..857C}, and 
\citet{2004A&A...419..249S,2005A&A...429.1007S} report 
brown dwarf periods that are very close to that limit. On the other hand, the 
Cody \& Hillenbrand sample contains only one period shorter than 14\,h, although
their sensitivity increases towards shorter periods. It remains to be 
confirmed whether some young brown dwarfs indeed rotate close to breakup 
or not. This is an important point to clarify, as rotation near breakup would 
be expected to have a substantial effect on interior structure, evolution, 
and angular momentum control for these objects. 

Several groups have searched for a relation between rotation and 
the presence of disks in the VLM regime, to probe for the
existence of disk braking. The evidence is ambiguous. The data presented in 
\citet{2010ApJS..191..389C} for the $\sigma$Ori cluster 'do not support a direct 
connection between rotation and the presence of a disk', but their sample 
may be too small for a definite statement. Moreover, their Fig. 15 does 
seem to show that fast rotators are mostly diskless, in line with 
the expectation for disk braking. For another small sample in the same 
cluster, \citet{2004A&A...419..249S} find tentative evidence for a  
a disk braking scenario.

For the much larger sample in the ONC, \cite{2010A&A...515A..13R} find 
that “objects with NIR excess tend to rotate slower than objects without NIR 
excess” in the mass regime between 0.075 and 0.4$\,M_{\odot}$. This is 
interpreted as evidence for disk braking. No such signature is seen in the 
substellar regime. One possible caveat here is that many brown dwarf disks 
show little or no excess emission in the NIR and require MIR data to be 
clearly detected. \citet{2005A&A...430.1005L} show clear signs of disk braking 
for the VLM stars in NGC2264, but also suggest that disk braking becomes less 
efficient in the VLM regime. Finally, for a diverse sample of young VLM 
stars and BDs, \citet{2005MmSAI..76..303M} show that objects “undergoing disk 
accretion are clearly seen to be preferentially slow rotators”.

Combining all these studies, we conclude that some form of disk braking seems 
to operate into the VLM regime and possibly in brown dwarfs as well. At least,
there is no strong evidence against this possibility. However, the 
exact mass dependence of this process remains to be clarified. This 
requires to either obtain sensitive mid-infrared data for a large sample of 
objects with known period or, conversely, periods for objects with known 
infrared SED.

\subsection{Summary}

The aforementioned findings can be summarised in a very concise form:
In brown dwarfs, some form of disk braking seems to be at work, but 
wind braking is not. 

\section{Disk evolution}
\label{s3}

Disks are relevant for a number of relatively obvious reasons. First, they are 
the matter and angular momentum resevoir of forming stars and thus may have an 
effect on the final properties of stars and brown dwarfs. Second, global disk 
properties could theoretically contain a fossil record of events in the 
earliest evolutionary stages that are difficult to observe. And third, disks are 
the birth places of planets, their evolution thus establishes the boundary 
conditions for the architecture of planetary systems. It is the latter aspect 
that will dominate the following 
discussion. Brown dwarf disks allow us to test planet formation in an extreme
parameter regime, which provides constraints on the robustness and ubiquity
of planet formation scenarios. 

Our interpretation of observations of brown dwarf disks is informed by
our understanding of disks around stars (for a recent review see 
\citet{2011ARA&A..49...67W}). In any given star forming regions,
three different types of dusty disks are observed: 1) strongly flared disks, 2) 
disks that are flatter presumably due to dust settling to the midplane, 3) 
disks with emission deficits in the near/mid-infrared, usually seen as signs 
for cleared out inner regions ('transition disks'). While it is tempting to 
see these three types as an evolutionary sequence, it is important to remember 
that we do not know whether all disks pass through these three stages, or not. 
Brown dwarf disks are useful environments to test processes like dust settling 
and inner disk clearing.

The first detections of dusty disks around brown dwarfs were published about
10-15 years ago, only very few years after the discovery of brown dwarfs
\citep{1998A&A...335..522C,2001ApJ...558L..51M,2001A&A...376L..22N,2002ApJ...571L.155T}.
The prevalence of disks down to masses of 0.01$\,M_{\odot}$ \citep{2002A&A...393..597N}
and the general similarity of stellar and substellar disks was established 
already in these early stages 
\citep[e.g.][]{2003AJ....126.1515J,2003ApJ...585..372L,2004ApJ...609L..33M}, 
albeit often only with anecdotal evidence. The
various infrared satellite missions of the recent years, in particular
Spitzer, Herschel, and WISE have for the first time made studies
of brown dwarf disks with meaningful samples feasible.

\subsection{Disk lifetimes}

In mid-infrared colours, there is usually a clear gap between objects with excess 
above the photosphere and those without, which means that the fraction of objects with
disk can be readily derived. Disk fractions are
robust as long as the underlying sample is sufficiently large and unbiased.
Using this technique, it is now ascertained that the disk fractions do not
drop substantially down to the lowest mass objects identified in star
forming regions \cite[e.g.][]{2008ApJ...672L..49S,2007A&A...472L...9Z}.

\begin{figure}[t!]
\resizebox{\hsize}{!}{\includegraphics[clip=true]{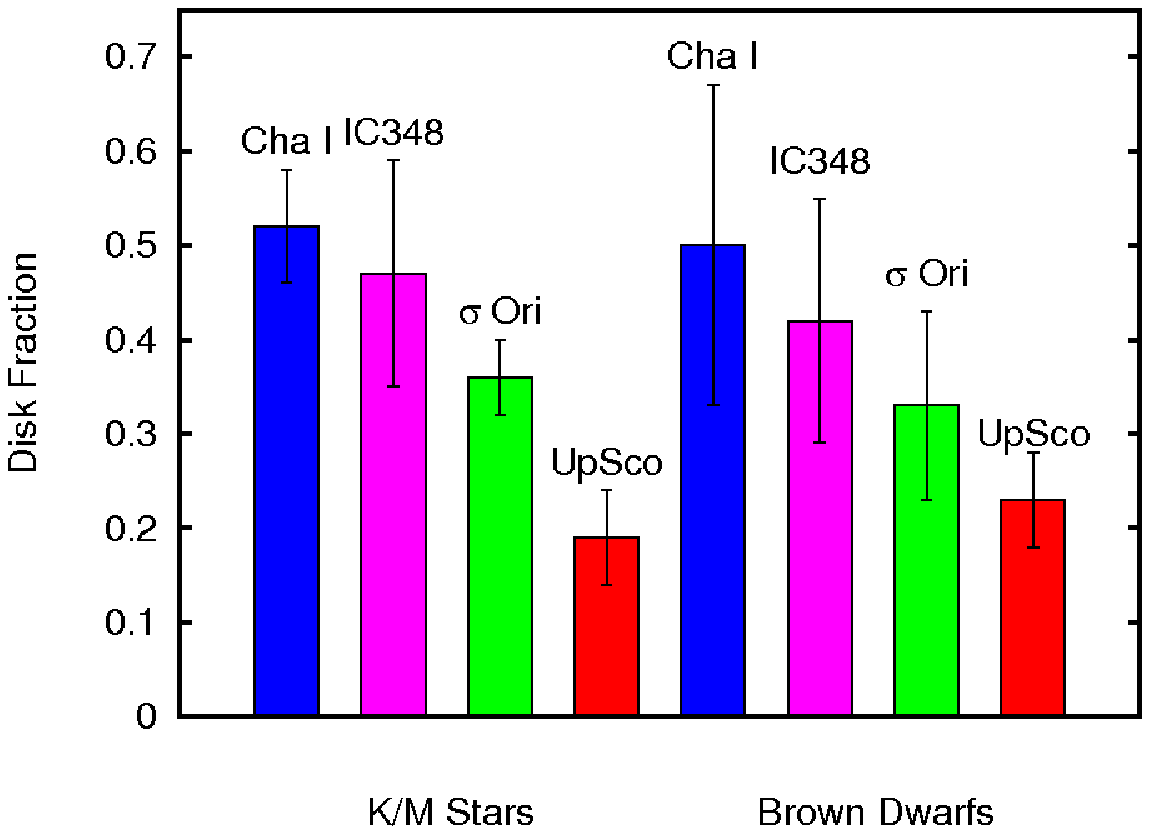}}
\caption{\footnotesize Disk fractions for low-mass stars and brown dwarfs
in various star forming regions, from \citet{2013MNRAS.429..903D}. The figure 
contains results from \citet{2005ApJ...631L..69L,2006AJ....131.1574L,2007ApJ...662.1067H,2007ApJ...670.1337D}.}
\label{f2}
\end{figure}

When the most recent studies with the largest samples are considered, the
disk fractions in the brown dwarf regime are consistent with those for low-mass
stars (K and M dwarfs) within the statistical limits. A good
example is the star forming region Upper Scorpius, with an age of 5-10\,Myr
an important benchmark test for the long-term evolution of disks. According
to \citet{2013MNRAS.429..903D}, its brown dwarf disk fraction is $23\pm 5$\% (Fig.
\ref{f2}), consistent with other recent estimates of this quantity
\citep{2013MNRAS.431.3222L,2012ApJ...758...31L}. For comparison, the disk 
fraction for K0-M5 stars is 19\% \citep{2006ApJ...651L..49C}. The disk fractions
of low-mass stars and brown dwarfs in the younger regions Chamaeleon-I, IC348, 
and $\sigma$\,Ori appear very similar as well (see Fig. \ref{f2}). This suggests that 
the overall disk lifetime in brown dwarf samples and the rate at which they 
lose their disks is not different from low-mass stars. 

The fraction of transition disks among brown
dwarf disks is again comparable to low-mass stars. For
example, \citet{2013MNRAS.429..903D} determined it as 19\% for Upper Scorpius, 
whereas values for low-mass stars are broadly in the 10-20\% range 
\citep{2009MNRAS.394L.141E,2010ApJ...708.1107M}. The errors in these numbers are 
large due to the small sample size, in addition, transition disks are a diverse
group of objects and the separation from 'normal' disks is often not unambiguous.
The brown dwarf transition disk fraction translates to clearing timescales in the order
of 0.4\,Myr or less, again comparable to low-mass stars, suggesting that the
clearing process is independent of the object mass and luminosity.
This could be a serious challenge for scenarios where 
photoevaporation driven by the ultraviolet radiation from the central object is 
the physical process responsible for the clearing \citep[e.g.][]{2006MNRAS.369..229A}. 
It follows that brown dwarf disks show a two-timescale evolution, with a 
long-term dissipiation and a much faster inner disk clearing, mirroring the 
behaviour of disks around low-mass stars.

\subsection{Dust settling}

A comparison of the SEDs of brown dwarf disks in diverse regions shows
that the median flux level drops with increasing age 
\citep{2012ApJ...744....6S}. More specifically, the fraction of highly flared 
disks drops with
age, and becomes negligible in the relatively old Upper Scorpius star forming
region \citep{2007ApJ...660.1517S}. This evolutionary behaviour is best explained
by dust settling to the midplane of the disk 
\citep[see discussion in][]{2009MNRAS.398..873S}, a consequence of grain growth.

More evidence for grain growth in brown dwarf disks comes from the analysis
of the silicate feature around 10$\,\mu m$. Brown dwarfs feature broad, 
flat silicate feature, more so than T Tauri stars and Herbig stars, indicating
an advanced stage of grain growth and grain removal from the upper layers
of the inner disk \citep{2005Sci...310..834A,2007ApJ...660.1517S}. This could 
simply be a result of the fact that this feature traces regions much closer to 
the object for cooler objects and the depletion of dust grains close to
the central source is faster due to the enhanced collision rates, although
other explanations are possible \citep{2009ApJ...696..143P}. 

Whether the dust settling inferred from the mid-infrared SEDs and thus
the growth of dust grains and the boundary conditions for the formation
of rocky planets is a function of the mass of the central object remains 
unclear. \citet{2010ApJ...720.1668S}
analyse mid-infrared SEDs and show that there is a significant difference in the
Spitzer/IRAC colour distributions of disks around low-mass and very-low mass
stars. In both mass bins, a degree of dust settling (i.e. flattened disks)
has to be assumed to explain the SEDs, however, as argued by 
\citet{2010ApJ...720.1668S}, 'relative to the disk structure predicted for 
flared disks, the required reduction in disk scale height is anti-correlated 
with the stellar mass'. In other words, disks around very low mass stars 
are flatter. This trend could continue into the brown dwarf regime.

A different approach with a different result was presented by 
\citet{2012A&A...539A...9M}. They model the SEDs of Herbig stars, T Tauri stars, 
and brown dwarfs in a
self-consistent way, and find that 'regions with the same temperature have a 
self-similar vertical structure independent of stellar mass'. However, regions at 
the same distance from the central object appear more settled in brown dwarfs,
due to their lower luminosities. Thus, the flatter disk structure of brown dwarfs
are more a result of their lower luminosities than different physical processes
in the disk evolution.
The turbulent mixing strength, parameterised
using the $\alpha$ prescription, is the same in all three samples. Thus, 
according to their analysis, disks around objects with a wide range of stellar
and substellar masses have self-similar structures and provide similar 
environments for the early stages of rocky planet formation.

\subsection{Submm/mm observations: the next frontier}

For many outstanding problems in this field the crucial spectral domain is the submm/mm
wavelength regime. At these wavelengths, the disks are mostly optically thin.
As a result, the emission directly traces the properties of the bulk of the dust
in the disk, including the total dust mass and the dust opacity.
Also, submm/mm observations provide
the best opportunity to resolve the disks and thus put limits on their physical
dimensions. 

The submm/mm domain remains the wavelength regime where our observational database is 
still very sparse. Only about a handful of brown dwarfs have been detected, using 
MAMBO-2 at IRAM, SCUBA and SCUBA-2 at JCMT as well as the SMA 
\citep{2003ApJ...593L..57K,2006ApJ...645.1498S,2008A&A...486..877B,2008ApJ...689L.141P,2013arXiv1305.6896M}.
These results tentatively indicate that the substellar disk
masses are around 0.5\%, roughly comparable to T Tauri stars. For comparison,
\citet{2012ApJ...744L...1H}
report relative disk masses of a few $10^{-5}$ to 
$10^{-2}$ times the central object masses, which could indicate a drop
in the relative disk masses compared with T Tauri stars. However, these
results are based on Herschel/PACS far-infrared fluxes, insensitive to
relatively cold and large dust grains and not in the optically thin
regime. At this point it might be useful to consider these values lower limits
when comparing with disk masses derived from submm/mm data.

The total radii of brown dwarf disks are very poorly constrained. 
Combining the result from the first resolved observations of a brown dwarf disk
\citep{2013ApJ...764L..27R} with some more indirect constraints 
\citep{2007ApJ...666.1219L,2006ApJ...645.1498S}, it seems that some 
brown dwarf disks have radii in the range of 10 to 40\,AU, about a factor of
ten smaller than the largest T Tauri disks.

\begin{figure}[t!]
\resizebox{\hsize}{!}{\includegraphics[clip=true]{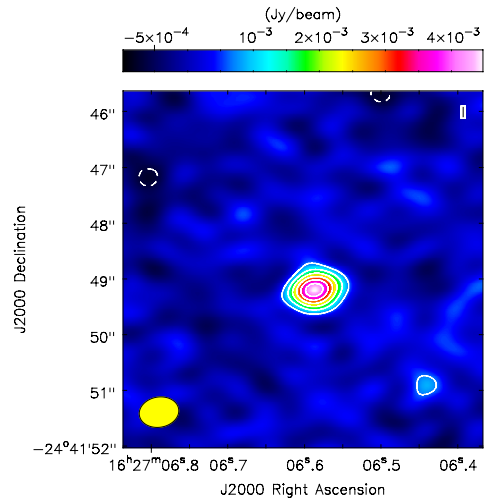}}
\caption{\footnotesize ALMA continuum map of RhoOph-102 at 0.89\,mm,
from \citet{2012ApJ...761L..20R}. White contour lines are drawn at 
-3,3,6,9,...,18$\sigma$, where $\sigma = 0.22$\,mJy/beam is the rms 
noise of the map. The filled yellow ellipse in the lower left corner
indicates the size of the synthesized beam.}
\label{f3}
\end{figure}

Recently, \citet{2012ApJ...761L..20R} reported the first observations of a brown
dwarf disk with the new submm/mm interferometer ALMA (see Fig. \ref{f3}). The 
main result from
this study is the detection of millimeter sized grains in the brown dwarf disk,
with a spectral slope that is comparable to T Tauri stars. 
This is 
so far the clearest demonstration that grain growth is a very robust 
process that occurs even in the low-density environments of brown dwarf
disks. The result also challenges the theory for grain growth. To stop the
strong radial drift, \citet{2013A&A...554A..95P} suggest the presence of 
extreme pressure bumps in the dust distribution to trap the particles. 
Whether such a scenario is pausible, remains to be explored (see also the
contribution by Ricci in these proceedings).

The 1mm detection by Ricci et al. took only 15 min total integration 
time with ALMA and yielded a flux uncertainty of 0.22\,mJy. For comparison,
the previous detections with IRAM/MAMBO-2 needed four times longer on-source
time for a noise level of $\sim 0.7$\,mJy \citep{2006ApJ...645.1498S}. Note 
that this was achieved with only the partially completed ALMA array (15 
antennaes). With the fully completed array, ALMA will fundamentally improve 
our knowledge. It will allow us to observe large samples in 
an efficient manner, push the sensitivity limits for disk masses to 
fractions of a Jupiter mass, study the gas component from the CO
emission, and also provides the opportunity to resolve large numbers of
brown dwarf disks. In combination with the completed surveys with 
Spitzer, WISE, and Herschel, the expected ALMA data will become a very
powerful probe of the disk evolution in substellar objects.

\subsection{Summary}

Brown dwarfs harbour disks
that are scaled down versions of T Tauri disks. All physical processes observed
in circumstellar disks are also found in their circumsubstellar siblings.
The global disk properties (mass and radius) seem to scale roughly with 
central object mass. Moreover, the timescales of the disk evolution are 
remarkably similar as well.

In terms of brown dwarf formation scenarios, this might be the expected outcome,
if brown dwarfs originate from the same processes that also form stars. However,
in terms of disk evolution models, the similarity of stellar and substellar
disks is somewhat surprising. Multiple processes that affect the
disk evolution are expected to be a strong function of the mass or luminosity of the 
central object or the mass of the disk. These processes, partly mentioned in
this review, include photoevaporation, disk
fragmentation, magneto-rotational-instability, and grain growth. 
Observations of brown dwarf disks can help to constrain all these aspects and 
may prove to be a crucial test case for our understanding of disk evolution.

Given the similarity of stellar and substellar disks, it is not too daring
to state that brown dwarfs are indeed likely to form their own miniature 
planetary systems, which could be scaled down versions of the diverse exoplanetary 
systems discovered
over the past decade. While giant planets will be very rare, a small population of 
Earth-sized planets could exist \citep{2007MNRAS.381.1597P}, in addition to smaller 
rocky planets and asteroid belts. The various issues for the habitability of these 
miniature planetary systems have been outlined by \citet{2013AsBio..13..279B}. While 
making a planet in orbit around a brown dwarf might be possible, living on such a 
planet seems enormously difficult.

\begin{acknowledgements}
I would like to thank Antonio Magazzu and Eduardo Martin as well
as all other members of the SOC and LOC for organising a stimulating
and, at the same time, relaxing conference. This contribution was
funded by the Science Foundation Ireland through grant no.
10/RFP/AST2780.
\end{acknowledgements}

\bibliographystyle{aa}

\end{document}